\documentclass[prb,aps,twocolumn,showpacs]{revtex4-2} %floatfix
\usepackage{amsmath,amssymb,amsthm}
\usepackage{graphicx}
\usepackage{subfigure}
\usepackage{amsmath}
\usepackage{amsfonts,amssymb}
\usepackage{bm}
\usepackage{float}
\usepackage{color}
\usepackage{sidecap}
\allowdisplaybreaks
\usepackage{soul}
\setstcolor{red}
\usepackage[normalem]{ulem}
\usepackage{hyperref}
\hypersetup{colorlinks=true,breaklinks,urlcolor=blue,linkcolor=blue,citecolor=blue}

\begin{document}

\title{Real spectra and phase transition of skin effect in nonreciprocal systems}
\author{Qi-Bo Zeng}
\email{zengqibo@cnu.edu.cn}
\affiliation{Department of Physics, Capital Normal University, Beijing 100048, China}

\author{Rong L\"u}
\affiliation{State Key Laboratory of Low-Dimensional Quantum Physics, Department of Physics, Tsinghua University, Beijing 100084, China}
\affiliation{Frontier Science Center for Quantum Information, Beijing 100084, China}

\begin{abstract}
We study the one-dimensional nonreciprocal lattices with real nearest neighboring hopping and find that the energy spectra under open boundary conditions can be entirely real or imaginary. We further investigate the spectral properties and the non-Hermitian skin effect in the one-dimensional mosaic lattices with real nonreciprocal hopping introduced at equally spaced sites. The eigenenergies of such lattices undergo a real-complex-imaginary or real-complex transition as the nonreciprocity varies. Moreover, the skin effect exhibits phase transitions depending on the period of the mosaic nonreciprocity. The bulk states are abruptly shifted from one end of the lattice to the opposite one by crossing the critical points, accompanied by the closing and reopening of point gaps in the spectra under periodic boundary conditions. The phase diagrams of the transition are presented and the critical boundaries are analytically determined. Our work unveils the intriguing properties of the energy spectrum and skin effect in non-Hermitian systems.
\end{abstract}
\maketitle
\date{today}

\section{Introduction}
The study on non-Hermitian systems has attracted much attention during the past few years~\cite{Cao2015RMP,Konotop2016RMP,Ganainy2018NatPhy,Ashida2020AiP,Bergholtz2021RMP}. Non-Hermiticity may appear due to the interactions between the systems and the external environment, which are commonly represented by physical gain and loss in the model Hamiltonian and can exist in both classical~\cite{Makris2008PRL,Klaiman2008PRL,Guo2009PRL,Ruter2010NatPhys,Lin2011PRL,Regensburger2012Nat,Feng2013NatMat,Peng2014NatPhys,Wiersig2014PRL,Hodaei2017Nat,Chen2017Nat} and quantum systems~\cite{Brody2012PRL,Lee2014PRX,Li2019NatCom,Kawabata2017PRL,Hamazaki2019PRL,Xiao2019PRL,Wu2019Science,Yamamoto2019PRL,Yamamoto2019PRL,Naghiloo2019NatPhys,Matsumoto2020PRL}. The energy spectra of non-Hermitian systems are normally complex but can be entirely real when $\mathcal{PT}$-symmetry~\cite{Bender1998PRL,Bender2002PRL,Bender2007RPP} or pseudo-Hermiticity~\cite{Mostafazadeh2002JMP,Mostafazadeh2010IJMMP,Moiseyev2011Book,Zeng2020PRB1,Kawabata2020PRR,Zeng2021arxiv} is respected. The existence of real spectra ensures the stabilities of non-Hermitian systems, which facilitates their applications in various fields.  

Recently, nonreciprocal systems have inspired considerable research interest~\cite{Lee2016PRL,Lieu2018PRB,Yin2018PRA}. The energy spectra of such systems are very sensitive to the change of boundary conditions~\cite{Xiong2018JPC}, which can be utilized to design new types of quantum sensors~\cite{Budich2020PRL,Koch2022PRR}. For systems with nonreciprocal hopping and under open boundary conditions (OBCs), the non-Hermitian skin effect (NHSE) will emerge with all the bulk states localized at the boundaries. The NHSE impacts a multitude of physical phenomena significantly~\cite{Shen2018PRL,Yao2018PRL1,Yao2018PRL2,Alvarez2018PRB,Alvarez2018EPJ,Lee2019PRB,Zhou2019PRB,Kawabata2019PRX,Okuma2020PRB,Xiao2020NatPhys,Yoshida2020PRR,Longhi2019PRR,Yi2020PRL}, such as the bulk-boundary correspondence of topological phases~\cite{Yao2018PRL1,Yao2018PRL2,Kunst2018PRL,Jin2019PRB,Yokomizo2019PRL,Herviou2019PRA,Borgnia2020PRL,Yang2020PRL2,Zirnstein2021PRL,Zhang2022arxiv} and Anderson localization phenomenon~\cite{Hatano1996PRL,Shnerb1998PRL,Gong2018PRX,Jiang2019PRB,Zeng2020PRR,Liu2021PRB1,Liu2021PRB2}. Though the NHSE is found to be connected to the point gap in the spectra under periodic boundary conditions (PBCs)~\cite{Okuma2020PRL,Zhang2020PRL}, how the behavior of NHSE is influenced by the variation of nonreciprocity is less studied. Recently, a new kind of lattice model called the mosaic lattice was proposed, where the modulations are imposed on equally spaced sites or hopping terms~\cite{Wang2020PRL,Zeng2021PRB}. The localization properties and topological phases are found to be influenced significantly in such systems. It will be interesting to check what will happen to the NHSE if nonreciprocal hopping are only periodically added to certain instead of all hopping terms.

In this paper, we investigate the one-dimensional (1D) nonreciprocal lattices with only real nearest neighboring hopping. We prove that as long as the products of each pair of the forward and backward hopping between the neighboring sites are positive (negative), these non-Hermitian matrices can be transformed into Hermitian (anti-Hermitian) ones through similarity transformations, indicating the existence of entirely real (imaginary) spectra. We further study the spectral properties and the NHSE in the 1D mosaic lattices with real nonreciprocal hopping introduced at equally spaced sites. We find that the OBC spectra of these systems undergo real-complex-imaginary or real-complex transitions as the strength of the nonreciprocity varies. To characterize the NHSE with bulk states localized at different ends of the 1D lattices, we introduce the directional inverse participation ratio (dIPR). Interestingly, we find that when the period of the nonreciprocity is not a multiple of that of the underlying reciprocal hopping, the NHSE will show a phase transition at nonzero nonreciprocity, where the bulk states are abruptly shifted from one end of the lattice to the opposite one. Such a phenomenon is accompanied by the closing and reopening of the point gaps in the PBC spectrum. We present the phase diagrams and analytically determine the critical boundaries of the phase transition. Our work reveals the interesting spectral properties and the phase transition of NHSE in 1D nonreciprocal lattices.

The rest of the paper is organized as follows. In Sec.~\ref{sec2} we study the energy spectra of the nearest-neighboring nonreciprocal lattices. Then we introduce the directional inverse participation ratio in Sec.~\ref{sec3} and use it to explore the phase transitions of NHSE in nonreciprocal dimer lattices in Sec.~\ref{sec4}. We further investigate the NHSE in trimer lattices and the off-diagonal Aubry-Andr\'e-Harper model with nonreciprocal hopping in Sec.~\ref{sec5}. The last section (Sec.~\ref{sec6}) is dedicated to a summary. 

\begin{figure}[t]
	\includegraphics[width=3.4in]{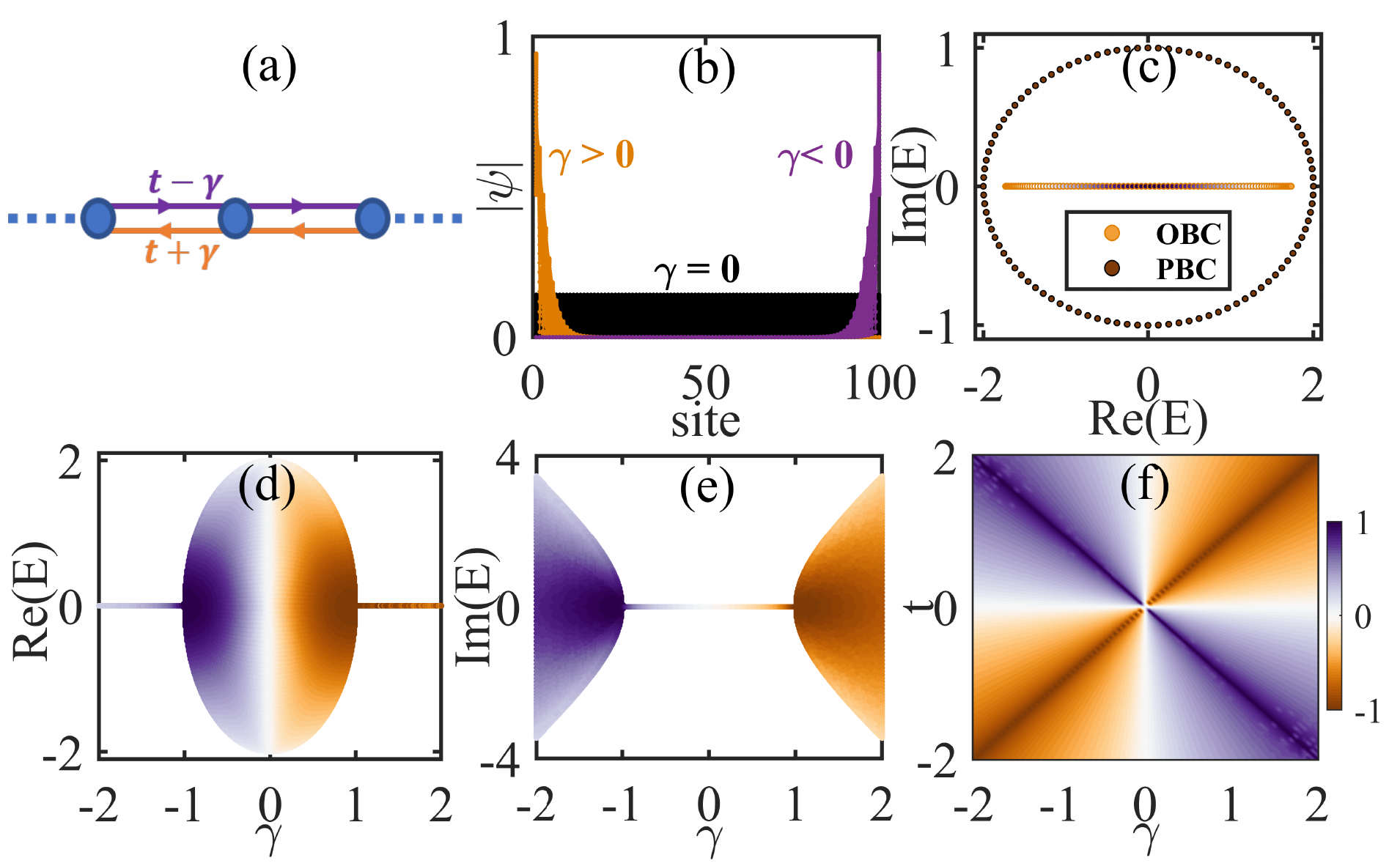}
	\caption{(Color online) (a) Schematic of the 1D Hatano-Nelson lattice model. (b) The bulk eigenstates at the left (brown) or right (purple) end of the lattice when $\gamma>0$ or $\gamma<0$ due to the NHSE. (c) The PBC and OBC spectrum of the model at $t=1$, $\gamma=0.5$. Panels (d) and (e) show the real and imaginary parts of the OBC spectrum as a function of $\gamma$. In (b)-(e), we set $t=1$. (f) Phase diagram of the NHSE for the HN model as a function of $t$ and $\gamma$. The colorbar indicates the value of directional MIPR in (f) and the dIPR of eigenstates in (d) and (e). The lattice size is $L=100$.}
	\label{fig1}
\end{figure}

\section{Energy spectra of the nearest-neighboring nonreciprocal lattices}\label{sec2}
One of the most famous nonreciprocal lattice models is the Hanato-Nelson (HN) model, where the hoppings between the neighboring sites are asymmetric, see Fig.~\ref{fig1}(a). The Hamiltonian is $H=\sum_j (t+\gamma)c_j^\dagger c_{j+1} + (t-\gamma)c_{j+1}^\dagger c_j$, with $c_j$ ($c_j^\dagger$) being the annihilation (creation) operator of spinless fermions at the $j$th site. The forward and backward hopping amplitudes are ($t-\gamma$) and ($t+\gamma$), respectively. If $t>0$ and $\gamma > 0$ (or $\gamma<0$), the eigenstates will be localized at the left end (or right end) of the 1D lattice [Fig.~\ref{fig1}(b)], which is the NHSE. Another important feature is the real spectrum in the HN model under open boundary conditions (OBCs). The corresponding spectrum under periodic boundary conditions (PBCs), however, is complex and forms a loop with a point gap, where a reference point inside the loop such as the origin of the complex energy plane cannot be crossed by the loop~\cite{Kawabata2019PRX}, as shown in Fig.~\ref{fig1}(c). The existence of point gap is the topological origin of the NHSE~\cite{Okuma2020PRL,Zhang2020PRL}. From the imaginary parts shown in Fig.~\ref{fig1}(e), we find that the OBC spectrum is always real when $|\gamma|<|t|$ for the HN model. 

Normally the real spectra in non-Hermitian systems are connected with the $\mathcal{PT}$-symmetry or pseudo-Hermiticity of the model Hamiltonian. Here we show that for the systems with only real and nearest neighboring hopping, we can prove that the model Hamiltonian is pseudo-Hermitian and thus can obtain an entirely real spectrum under certain conditions. Specifically, for a general 1D lattice with nearest-neighboring hopping, the Hamiltonian matrix is a tridiagonal matrix and can be represented as 
\begin{equation}\label{H}
	H = \left(
	\begin{array}{ccccc}
		V_1 & t_1 & 0 & \cdots & 0 \\
		t^\prime_1 & V_2 & t_2 & \cdots & 0 \\
		0 & t^\prime_2 & V_3 & t_3 & \cdots \\
		\vdots & \vdots & \vdots & \ddots & \vdots \\
		0 & 0 & 0 & \cdots & V_L
	\end{array}
	\right),
\end{equation}
where $V_j$ along the main diagonal is the onsite potential on the $j$th site and $t_j$ ($t^\prime_j$) is the backward (forward) hopping amplitude between the $j$th and the $(j+1)$th lattice sites. $L$ is the size of the lattice. All the elements in the matrix are real and we have $t_j \neq t^\prime _j$, implying that the matrix is non-Hermitian. Such a tridiagonal matrix represents the 1D nonreciprocal lattices with asymmetric hopping between the nearest neighboring sites under open boundary conditions. If we further assume that $t_j t^\prime_j>0$, then we can transform the non-Hermitian matrix $H$ into a Hermitian one through a similarity transformation. To do so, we construct a diagonal matrix as
\begin{equation}\label{D}
	D = \left(
	\begin{array}{ccccc}
		d_1 & 0 & 0 & \cdots & 0 \\
		0 & d_2 & 0 & \cdots & 0 \\
		0 & 0 & d_3 & 0 & \cdots \\
		\vdots & \vdots & \vdots & \ddots & \vdots \\
		0 & 0 & 0 & \cdots & d_L \\
	\end{array}
	\right),
\end{equation}
with the diagonal element $d_j$ defined as
\begin{equation}\label{dj}
	d_j = \left\{
	\begin{array}{c}
		1, \qquad j = 1 \\
		\sqrt{\frac{t^\prime_{j-1} t^\prime_{j-2} \cdots t^\prime_1}{t_{j-1}t_{j-2} \cdots t_1}}, \qquad j = 2,3,\cdots,L
	\end{array}
	\right.
\end{equation}
Then the transformation leads to the following Hermitian matrix 
\begin{widetext}
\begin{equation}
	h = D^{-1}HD = \left(
	\begin{array}{ccccc}
		V_1 & sgn(t_1)\sqrt{t_1 t^\prime_1} & 0 & \cdots & 0 \\
		sgn(t_1)\sqrt{t_1 t^\prime_1} & V_2 & sgn(t_2)\sqrt{t_2 t^\prime_2} & \cdots & 0 \\
		0 & sgn(t_2)\sqrt{t_2 t^\prime_2} & V_3 & sgn(t_3)\sqrt{t_3 t^\prime_3} & \cdots \\
		\vdots & \vdots & \vdots & \ddots & \vdots \\
		0 & 0 & 0 & \cdots & V_L \\
	\end{array}
	\right).
\end{equation}
\end{widetext}
Since the spectrum of the Hermitian matrix $h$ are real, the Hamiltonian $H$ will also host real spectrum as long as these two can be transformed into each other by the similarity transformation shown above. So the condition for the existence of real spectra for the tridiagonal Hamiltonian matrix $H$ is $t_j t^\prime_j>0$. In fact, the similarity transformation actually implies that the Hamiltonian is pseudo-Hermitian, which can be proved as follows
\begin{equation}
	D^2 H^\dagger D^{-2} =D^2 (D h D^{-1})^\dagger  D^{-2}= D h D^{-1} = H,
\end{equation}
which means $H^\dagger = \eta^{-1} H \eta$ with $\eta=D^2$. Thus, the Hamiltonian is pseudo-Hermitian, as originally defined by Mostafazadeh~\cite{Mostafazadeh2002JMP}.

Since the similarity transformation does not depend on the specific formula of the parameters, the conclusion can be applied to very general situations. For instance, the method can be used to analyze the real-complex transition in the spectra of the quasireciprocal lattices with random nonreciprocal hopping~\cite{Zeng2021arxiv}.

The above similarity transformation can also be used to prove the existence of imaginary spectra. Suppose that now the elements on the main diagonal are imaginary, and the Hamiltonian becomes
\begin{equation}
	H_I = \left(
	\begin{array}{ccccc}
		iV_1 & t_1 & 0 & \cdots & 0 \\
		t^\prime_1 & iV_2 & t_2 & \cdots & 0 \\
		0 & t^\prime_2 & iV_3 & t_3 & \cdots \\
		\vdots & \vdots & \vdots & \ddots & \vdots \\
		0 & 0 & 0 & \cdots & iV_L
	\end{array}
	\right),
\end{equation}
with $V_j$, $t_j$, and $t^\prime_j$ being real numbers. The diagonal matrix defined in Eqs.~(\ref{D}) and (\ref{dj}) can also be used to transform $H_I$ as
\begin{equation}
	h_I = D^{-1}H_I D = \left(
	\begin{array}{ccccc}
		iV_1 & \sqrt{t_1 t^\prime_1} & 0 & \cdots & 0 \\
		\sqrt{t_1 t^\prime_1} & iV_2 & \sqrt{t_2 t^\prime_2} & \cdots & 0 \\
		0 & \sqrt{t_2 t^\prime_2} & iV_3 & \sqrt{t_3 t^\prime_3} & \cdots \\
		\vdots & \vdots & \vdots & \ddots & \vdots \\
		0 & 0 & 0 & \cdots & iV_L \\
	\end{array}
	\right).
\end{equation}
Then if the products of $t_j t^\prime_j$ ($j=1,2,\cdots,L$) are all negative, i.e., $t_j t^\prime_j<0$, we have
\begin{equation}
	h_I = i \left(
	\begin{array}{ccccc}
		V_1 & \sqrt{|t_1 t^\prime_1|} & 0 & \cdots & 0 \\
		\sqrt{|t_1 t^\prime_1|} & V_2 & \sqrt{|t_2 t^\prime_2|} & \cdots & 0 \\
		0 & \sqrt{|t_2 t^\prime_2|} & V_3 & \sqrt{|t_3 t^\prime_3|} & \cdots \\
		\vdots & \vdots & \vdots & \ddots & \vdots \\
		0 & 0 & 0 & \cdots & V_L \\
	\end{array}
	\right) = iH_R,
\end{equation}
where $H_R$ is a Hermitian matrix with real spectra. Thus $h_I$ is anti-Hermitian; the spectra of $h_I$ and $H_I$ are entirely imaginary.

For the 1D Hanato-Nelson (HN) lattices, we have $V_j=0$, $t_j=t+\gamma$, and $t^\prime_j = t-\gamma$. If $(t+\gamma)(t-\gamma)>0$, which corresponds to $|\gamma|<|t|$, then the energy spectrum is entirely real. If $(t+\gamma)(t-\gamma)<0$, which corresponds to $|\gamma|>|t|$, then the energy spectrum is entirely imaginary. So the spectrum of the HN model undergoes a real-imaginary transition as we tune the nonreciprocity $\gamma$. The critical point for the transition is $\gamma_c = |t|$, which are consistent with the numerical results shown in Fig.~\ref{fig1}.   

\begin{figure}[t]
	\includegraphics[width=3.0in]{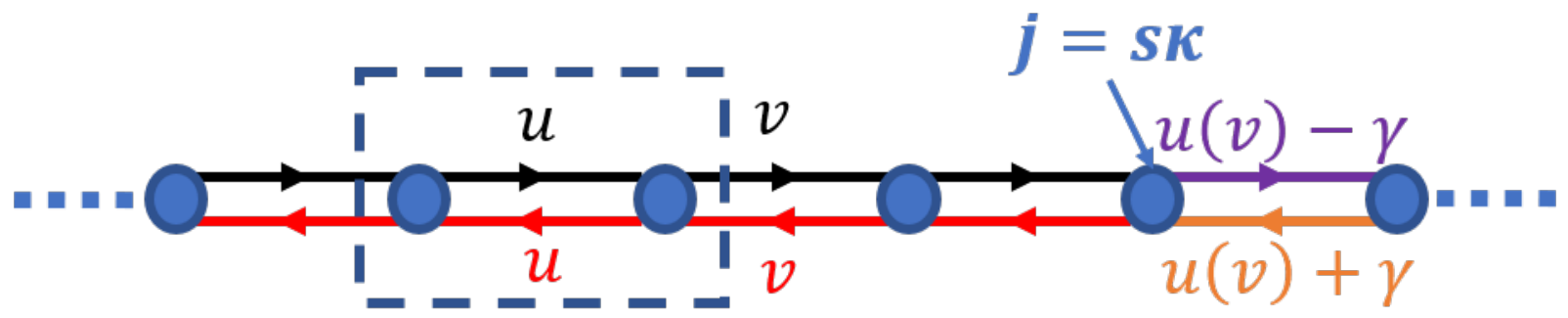}
	\caption{(Color online) Structure of the 1D mosaic nonreciprocal dimer lattices. The dashed rectangle indicates the unit cell containing two sites with intracell hopping $u$ and intercell hopping $v$. The nonreciprocal hopping is added to the hopping terms between the $j$th and $(j+1)th$ site with $j=s\kappa$. Depending on the value of $\kappa$, the nonreciprocal hopping amplitudes will be $u \pm \gamma$ or $v \pm \gamma$.}
	\label{fig2}
\end{figure}

\begin{figure}[t]
	\includegraphics[width=3.4in]{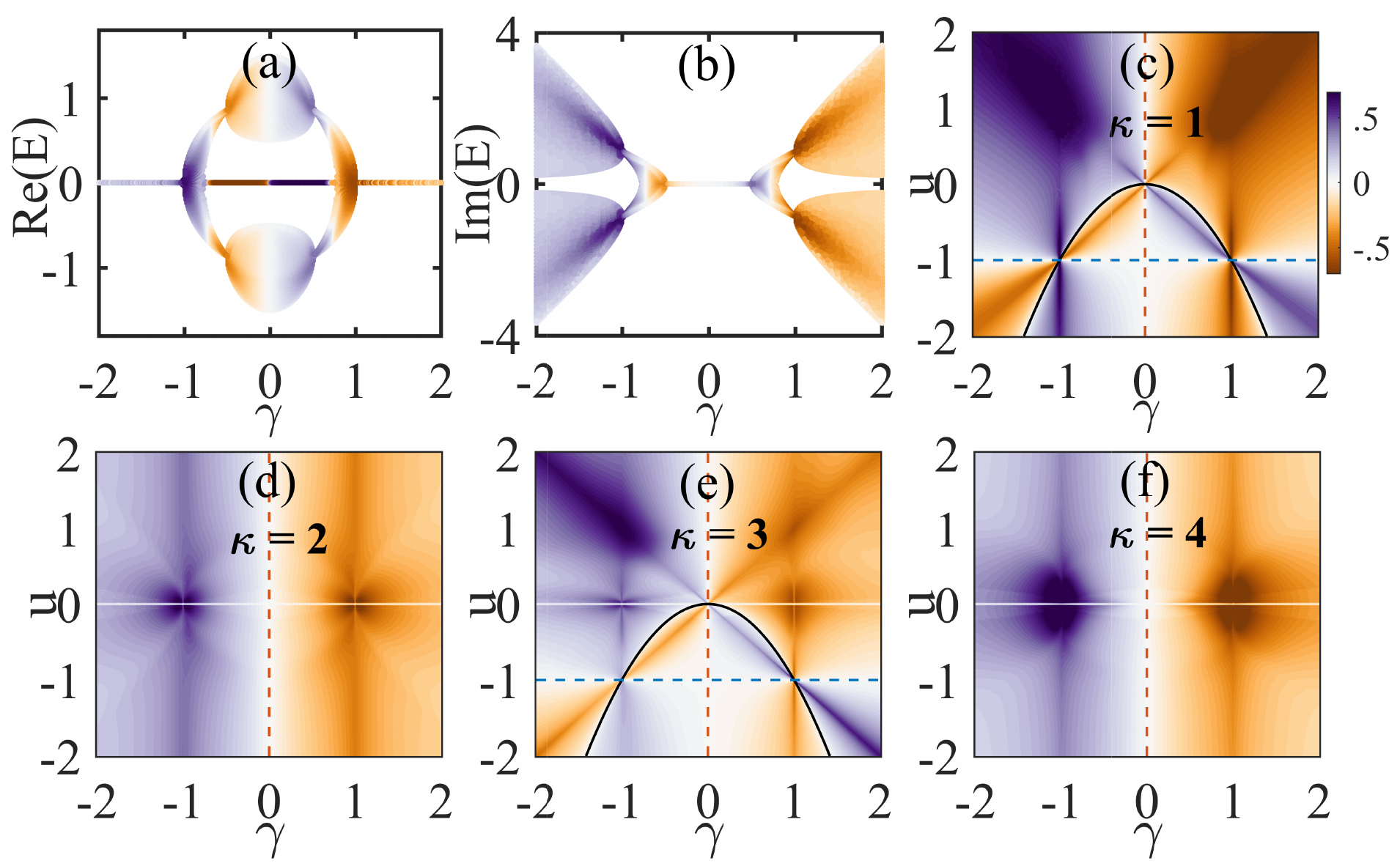}
	\caption{(Color online) Real (a) and imaginary parts (b) of the spectrum for the 1D mosaic nonreciprocal dimer lattice with $u=-0.5$, $\kappa=1$. Panels (c)-(f) show the phase diagrams of NHSE for the dimer lattices with $\kappa=1$, $2$, $3$, and $4$, respectively. The solid and dashed lines represent the critical lines of the phase transition. Other parameters: $v = 1$; $L=100$.}
	\label{fig3}
\end{figure}

As to the 1D mosaic nonreciprocal lattice models we will study in the following, the Hamiltonian matrices under OBCs are also tridiagonal with zero diagonal elements, i.e., $V_j=0$ for all $j$s. When $j=s\kappa$, the nearest-neighboring hopping amplitudes are of the form $t_j \pm \gamma$. To get the real spectrum, we must have $(t_j+\gamma)(t_j-\gamma)>0$, so the regimes for real spectrum is $|\gamma|<|t_j|$ with $t_j$ being the smallest hopping amplitude in the reciprocal lattices. Taking the mosaic nonreciprocal dimer lattices shown in Fig.~\ref{fig2} as an example, the model Hamiltonian is
\begin{align}\label{Hd}
	H_{dimer} =& \sum_{j=1}^{N_c} \left[ u c_{j,A}^\dagger c_{j,B} + v c_{j,B}^\dagger c_{j+1,A} + \text{H.c.} \right] \\ \notag
	&+ \sum_{s} \left[ \gamma c_{s\kappa}^\dagger c_{s\kappa+1} - \gamma c_{s\kappa+1}^\dagger c_{s\kappa} \right],
\end{align} 
with $N_c$ being the number of unit cell and system size being $L=2N_c$. When $\gamma=0$, each unit cell contains two sites with the intracell and intercell hopping being $u$ and $v$, respectively. The nonreciprocal hopping is only added to hopping terms between the $(s\kappa)$th and the $(s\kappa+1)$th site with $\kappa$ and $s$ both being positive integers. So the nonreciprocity appears every $\kappa$ site. When $\kappa=1$, the corresponding Hamiltonian matrix is 
\begin{equation}
	H_{dimer} = \left(
	\begin{array}{ccccc}
		0 & u+\gamma & 0 & \cdots & 0 \\
		u-\gamma & 0 & v+\gamma & \cdots & 0 \\
		0 & v-\gamma & 0 & u+\gamma & \cdots \\
		\vdots & \vdots & \vdots & \ddots & \vdots \\
		0 & 0 & 0 & \cdots & 0
	\end{array}
	\right).
\end{equation}
The condition for the real spectrum is 
\begin{equation}
	\left\{
	\begin{array}{c}
		(u+\gamma)(u-\gamma)>0, \\
		(v+\gamma)(v-\gamma)>0,
	\end{array}
	\right.
\end{equation}
so the regime for the real spectrum is $|\gamma| < min(|u|,|v|)$. On the other hand, the condition for the imaginary spectrum is 
\begin{equation}
	\left\{
	\begin{array}{c}
		(u+\gamma)(u-\gamma)<0, \\
		(v+\gamma)(v-\gamma)<0,
	\end{array}
	\right.
\end{equation}
so the regime for the imaginary spectrum is $|\gamma| > max(|u|,|v|)$. For $min(|u|,|v|)<|\gamma|<max(|u|,|v|)$, the spectrum is complex. The energy spectra of the 1D nonreciprocal dimer lattices thus undergo a real-complex-imaginary transition as the strength of nonreciprocity $\gamma$ grows. The critical point for the real-complex transition is $|\gamma_{c1}|=min(|u|,|v|)$, while the critical point for the complex-imaginary transition is $|\gamma_{c2}|=max(|u|,|v|)$.  In Fig.~\ref{fig3}(a) and (b), we present the real and imaginary parts of the energy spectrum for the nonreciprocal dimer lattice with $u=-0.5$, $v=1$, and $\kappa=1$. The OBC spectrum is entirely real when $|\gamma|<0.5$ and entirely imaginary when $|\gamma|>1$. The spectrum thus shows a real-complex-imaginary transition as we tune the strength of nonreciprocity. These numerical results are consistent with the above analytical conclusions.

For a general mosaic nonreciprocal lattice without onsite potentials, the energy spectra will always undergo a real-complex-imaginary (real-imaginary for HN model) or real-complex transition as $\gamma$ increases. In such lattices, the nonreciprocity only appears at equally spaced sites and is determined by the value of $\kappa$. If $\kappa=1$, nonreciprocity exists in every hopping term and the hopping amplitudes can be expressed as $t_j=t_{j0}+\gamma$ and $t_j^\prime=t_{j0}-\gamma$. The product of each pair of forward and backward hopping is
\begin{equation}
	t_j t_j^\prime=(t_{j0}+\gamma)(t_{j0}-\gamma)=(t_{j0}^2 - \gamma^2), \qquad j=1,2,\cdots,L.
\end{equation}
So when $|\gamma|<min(|t_{j0}|)$, the spectra will be entirely real; while if $|\gamma|>max(|t_{j0}|)$, the spectra will be entirely imaginary. If $min(|t_{j0}|)<|\gamma|<max(|t_{j0}|)$, the spectra are complex. Thus the spectra undergo a real-complex-imaginary transition for lattices with $\kappa=1$ when the strength of nonreciprocity increases. The HN lattice is a special case with $t_{j0}=t$. If $\kappa \neq 1$, then some hopping terms will be reciprocal and some of the products of $t_j t_j^\prime$ will not change their signs since they are not dependent on the value $\gamma$. When $|\gamma|$ is small, we have real spectra. But if $|\gamma|$ becomes larger, some of the positive products of $t_j t_j^\prime$ will become negative while others will remain unchanged. So the similarity transformation will not hold and the spectra become complex. However, the products of $t_j t_j^\prime$ cannot be negative for all hopping terms, so we cannot get entirely imaginary spectra. Thus we only have real-complex transitions in the mosaic nonreciprocal lattices with $\kappa \neq 1$.

From the above discussions, we conclude that for the system with real nearest-neighboring nonreciprocal hopping, as long as the products of each pair of the forward and backward hopping are positive, we can transform the non-Hermitian Hamiltonian matrix into a Hermitian one through a similarity transformation, indicating the existence of real spectra. The existence of such similarity transformation actually implies the pseudo-Hermiticity of the model Hamiltonian. On the other hand, if the products of $t_j t^\prime_j$ for all the hopping terms are negative, then the model Hamiltonian can be transformed into an anti-Hermitian matrix, indicating the existence of imaginary spectra. When the sign of $t_j t^\prime_j$ changes for some but not all $j$s, the similarity transformation will break down, and the energy spectrum becomes complex. The critical point for such transition satisfies $t_j t^\prime_j=0$. In addition, if we further introduce on site potentials into the lattice, then the condition for the real (or imaginary) spectrum still holds when the on site potentials are purely real (or imaginary). Notice that the similarity transformation is only established for the systems with nearest nonreciprocal hopping; if the nonreciprocal hopping terms become long range, the OBC spectra will become complex~\cite{Zeng2022arxiv}. 

\section{NHSE and directional inverse participation ratio}\label{sec3}
To characterize the localization of the states we can use the inverse participation ratio (IPR), which is defined as $\text{IPR}=\sum_{j=1}^L {|\Psi_{n,j}|^4}/(\langle \Psi_n | \Psi_n \rangle)^2$ and has been extensively used in describing the Anderson localization phenomenon. Here $\Psi_n$ is the right eigenstate with component $\Psi_{n,j}$ and satisfies the Schr\"odinger equation $H \Psi_n = E_n \Psi_n$, with $H$ being the model Hamiltonian and $E_n$ the $n$th eigenenergy. For extended states, the IPR is close to 0; while for localized states, the IPR value is of the order $O(1)$. As to the NHSE, the bulk states can be localized at either end of the 1D lattice, it will be beneficial to distinguish those states by IPR. Hence we introduce the directional IPR (dIPR) as
\begin{equation}
	\text{dIPR} (\Psi_n) = \mathcal{P}(\Psi_n) \sum_{j=1}^L \frac{|\Psi_{n,j}|^4}{(\langle \Psi_n | \Psi_n \rangle)^2}, 
\end{equation}
with $\mathcal{P}(\Psi_n)$ defined as 
\begin{equation}
	\mathcal{P}(\Psi_n) = sgn \left[ \sum_{j=1}^L \left(  j- \frac{L}{2} - \delta \right) |\Psi_{n,j}| \right].
\end{equation}
Here $\delta$ is a positive value and is normally set to be $0<\delta<0.5$. $sgn(x)$ takes the sign of the argument, which is positive for $x>0$ and negative for $x<0$. $\mathcal{P}(\Psi_n)$ extracts the information of whether the eigenstate $\Psi_n$ is more localized at the left or the right half of the lattice. The dIPR is positive when $\Psi_n$ is localized at the right end, but is negative when $\Psi_n$ is localized at the left end. If $\text{dIPR} \rightarrow 0$, the state is extended. In Fig.~\ref{fig1}(d) and (e), we present the energy spectrum of the HN model featured by the dIPR. The dIPR behaves consistently with the distribution of bulk states shown in Fig.~\ref{fig1}(b). The bright white line indicates that the states are extended at $\gamma=0$. We further present the phase diagram of the NHSE in Fig.~\ref{fig1}(f). Here the colorbar corresponds to the average values of the dIPR, which is called the directional MIPR (dMIPR) and is calculated as 
\begin{equation}
	\text{dMIPR}=\frac{1}{L}\sum_{n=1}^{L} \text{dIPR}(\Psi_n)
\end{equation}
We find that the behavior of NHSE depends on the signs of both $t$ and $\gamma$. Besides, along the $t=\pm \gamma$ lines, the dMIPR is quite large, implying that the bulk states are extremely localized at the boundaries. This is because when $t=\pm \gamma$, the hopping is finite in one direction but zero in the other and the states can only be transported to one end. So the dIPR and dMIPR are quite useful in describing the behavior of NHSE in 1D lattices. 

\section{NHSE in mosaic nonreciprocal dimer lattices}\label{sec4}
Now we turn to the non-Hermitian skin effect in the 1D dimer lattices where the hopping amplitudes are staggered; see Fig.~\ref{fig2} and the Hamiltonian in Eq.~(\ref{Hd}).  If we set $u=v$, the model reduces to the HN model with nonreciprocal hopping imposed at equally spaced sites. The phase diagram of such mosaic HN models is the same as that of the regular HN model shown in Fig. 1(f).

From the variation of the dIPR shown in Figs.~\ref{fig3}(a) and \ref{fig3}(b), we can find that as $|\gamma|$ increases, the bulk states will be shifted from one end to the other. The critical value is $\gamma_c=\pm 1/\sqrt{2}$ other than $0$. For instance, when $-1/\sqrt{2}<\gamma<0$, the bulk states are localized at the left end of the lattice with negative dIPR. At $\gamma=-1/\sqrt{2}$, the dIPR becomes zero, implying the states become extended. If $\gamma<-1/\sqrt{2}$, all the bulk states are localized at the right end with positive dIPR values. We give the phase diagram by calculating the dMIPR values for the dimer lattices with $\kappa=1$ in Fig.~\ref{fig3}(c). When $u>0$, the bulk states will always be localized at the same side if the sign of $\gamma$ remains unchanged. However, when $u<0$, then the phase diagram will be divided into several regimes. The solid black line is the critical line determined by the equation $u=-\gamma^2/v$, while the two dashed lines correspond to the $\gamma=0$ and $u=-v$, respectively. The eigenstates corresponding to these critical lines are extended. The bulk states will be abruptly shifted from one end to the opposite one by crossing the critical lines. Interestingly, the same diagram is found for the mosaic lattice with $\kappa=3$, see Fig.~\ref{fig3}(e), while for the mosaic lattices with $\kappa=2$ and $4$, the NHSE only depends on the sign of $\gamma$, as shown in Figs.~\ref{fig3}(d) and \ref{fig3}(f). In fact, the phase diagram is only determined by whether $\kappa$ is an even or odd integer. Notice that in Figs.~\ref{fig3}(d)-\ref{fig3}(f), there is a white line at $u=0$, implying that the dMIPR values are close to $0$, but it is not a critical line for the transition. The eigenstates there are localized instead of extended. Since the lattice is disconnected at certain sites when $u=0$, the eigenstates are localized in the bulk. Normally the localized states are distributed evenly in the left and right half lattice, leading to a vanishing dMIPR.

\begin{figure}[t]
	\includegraphics[width=3.4in]{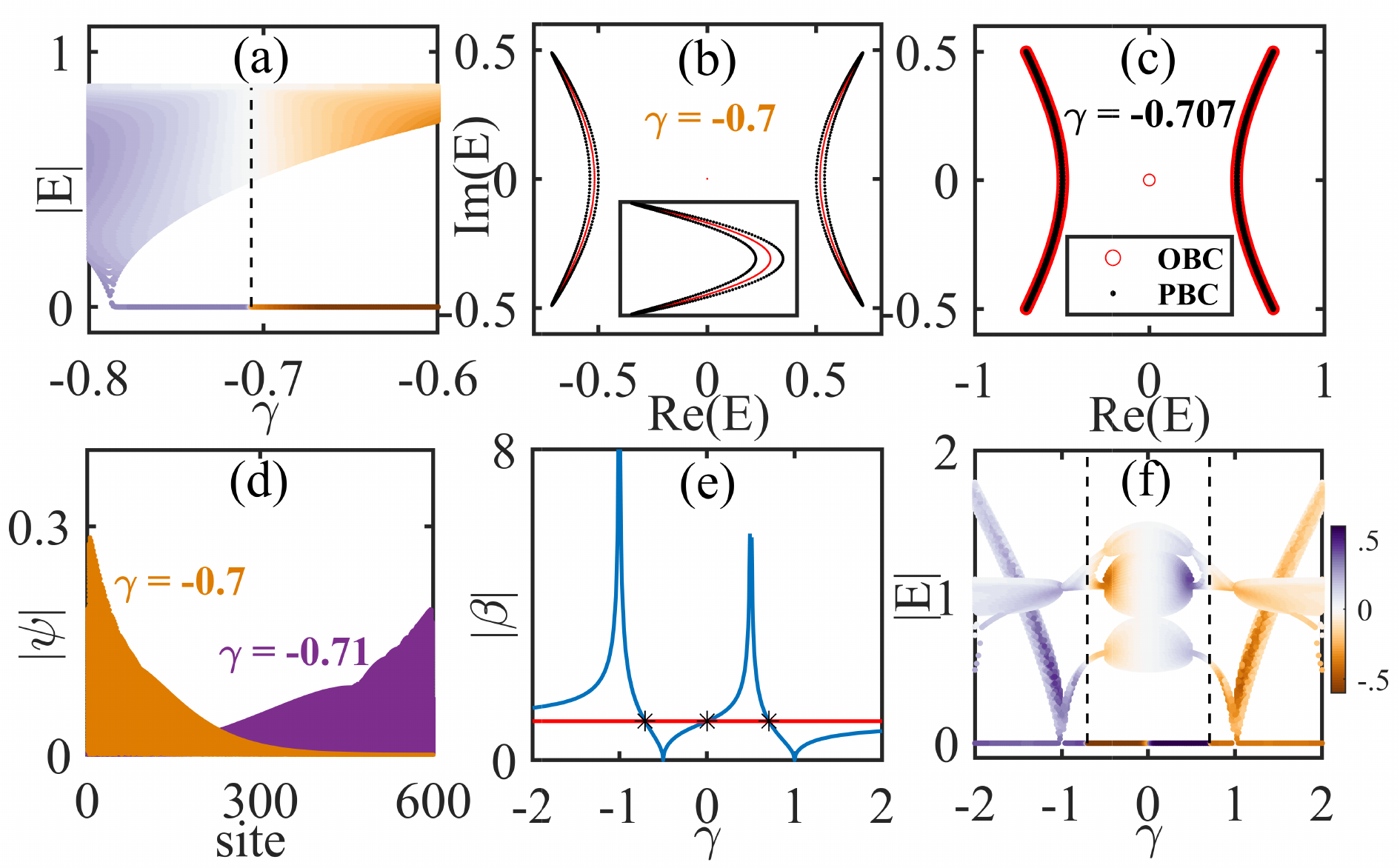}
	\caption{(Color online) (a) Eigenenergies $|E|$ of the nonreciprocal dimer lattices with $\kappa=1$. The black dashed line at $\gamma_c=-1/\sqrt{2}$ separates the phase with bulk states localized at opposite ends of the lattice. (b)(c) The PBC (black dots) and OBC (red dots) eigenenergies of the system at $\gamma=-0.7$ and $-0.707$ with a zooming up shown in the inset. Panel (d) shows the distribution of the bulk eigenstates  at $\gamma=-0.7$ and $\gamma=-0.71$, respectively. (e) The variation of $|\beta|$ as a function of $\gamma$. The stars represent the points satisfying $|\beta|=1$, which are located at $\gamma=0$ and $\pm 1/\sqrt{2}$. Other parameters: $u=-0.5$; $v = 1$; $L=600$. Panel (f) shows the spectrum for the dimer lattices with $\kappa=3$. The two black lines are the critical nonreciprocity at $\gamma=\pm 1/\sqrt{2}$. Here the lattice size is chosen to be $L=120$.}
	\label{fig4}
\end{figure}

It is known that the NHSE is connected to the point gap in the PBC spectra~\cite{Okuma2020PRL,Zhang2020PRL}. In Fig.~\ref{fig4}(a), we present the enlargement of the spectrum for the dimer lattice with $u=-0.5$, $v=1$, and $\kappa=1$. The dashed line corresponds to the critical value $\gamma_c=-1/\sqrt{2}$. Figure~\ref{fig4}(b) shows the PBC and OBC spectra at $\gamma=-0.7$. We can see that the PBC eigenenergies form loops, indicating the existence of point gaps. However, no loop structure is found in the OBC spectrum. When $\gamma=-1/\sqrt{2}$, the PBC and OBC spectrum become identical except for the topological zero modes, and the point gaps are closed, as shown in Fig.~\ref{fig4}(c). When $\gamma$ deviates from the critical value, the point gap will reopen and the bulk states are shifted to the boundaries, see Fig.~\ref{fig4}(d). The closing and reopening of the point gap signify a phase transition with the direction of NHSE reversed. So at the critical lines, the point gap in the PBC spectra should vanish. By diagonalizing the Bloch Hamiltonian, we have 
\begin{widetext}
\begin{equation}
	E_\pm = \pm \sqrt{u^2 + v^2 - 2\gamma^2 + 2uv \cos k + 2\gamma^2 \cos k + i 2\gamma (u+v) \sin k},
\end{equation}
\end{widetext}
with $k\in [0,2\pi)$ being the momentum. When the real or imaginary part of the energy is independent of $k$, the eigenenergies will not form loops in the complex energy plane, and there will be no point gap in the PBC spectrum. Then we have 
\begin{align}
	2uv \cos k + 2 \gamma^2 \cos k = 0 & \rightarrow u = -\frac{\gamma^2}{v}; \\
	2\gamma (u+v) \sin k = 0 & \rightarrow \gamma=0 \quad or \quad u = -v;
\end{align}
which are exactly the critical lines separating the regimes with states localized at different ends shown in Fig.~\ref{fig3}.

The phase transition of NHSE can also be explained by using the non-Bloch band theory~\cite{Yao2018PRL1,Yao2018PRL2,Yokomizo2019PRL}. The non-Bloch band theory is introduced to resolve the discrepancy between the eigenenergy spectra under periodic and open boundary conditions and restore the bulk-boundary correspondence in the non-Hermitian topological systems. For a periodic lattice with $q$ sites in each unit cell without onsite potentials, the backward hopping amplitudes between the nearest neighboring sites are represented by ${t_1, t_2,\cdots,t_q}$ and the corresponding forward hopping amplitudes are ${t^\prime_1,t^\prime_2,\cdots,t^\prime_q}$. The generalized Bloch Hamiltonian is defined as 
\begin{equation}
	H(\beta)= \left(
	\begin{array}{ccccc}
		0 & t_1 & 0 \cdots & t^\prime_q \beta^{-1} \\
		t^\prime_1 & 0 & t_2 & \cdots & 0 \\
		0 & t^\prime_2 & 0 & \cdots & 0 \\
		\vdots & \vdots & \vdots & \ddots & \vdots \\
		t_q \beta & 0 & \cdots & \cdots & 0 
	\end{array}
	\right),
\end{equation}
where $\beta=r e^{ik}$ is used to replace the $e^{ik}$ in the conventional Bloch Hamiltonian and is defined as
\begin{equation}
	|\beta|=r=\sqrt{ \left| \frac{t^\prime_1 t^\prime_2 \cdots t^\prime_q}{t_1 t_2 \cdots t_q} \right|}.
\end{equation}
For systems with $|\beta|<1$ ($|\beta|>1$), the bulk states will be localized at the left (right) end of the 1D lattice. If $|\beta|=1$, the bulk states are extended. We can determine the critical points for the phase transition where the bulk states are shifted from one end to the other by solving the equation $|\beta|=1$.

For the dimer lattices with nonreciprocity added on each hopping term, i.e., $\kappa=1$, we have two sites in each unit cell and the Bloch Hamiltonian is 
\begin{equation}
	h(k) = \left( 
	\begin{array}{cc}
		0 & (u + \gamma) + (v - \gamma) e^{-ik} \\
		(u - \gamma) + (v + \gamma) e^{ik} & 0 
	\end{array}
	\right).
\end{equation}
The generalized Bloch Hamiltonian $h(k)$ is obtained by replacing the $e^{\pm ik}$ with $\beta^{\pm 1}$, where 
\begin{equation}\label{betaD}
	|\beta| = \sqrt{\left| \frac{(u-\gamma)(v-\gamma)}{(u+\gamma)(v+\gamma)} \right| }.
\end{equation}
Solving the equation $|\beta|=1$, we have 
\begin{equation}
	\gamma = 0; \qquad u = -v; \qquad u = -\frac{\gamma^2}{v}.
\end{equation}
$\gamma=0$ is the trivial solution corresponding to the Hermitian systems without nonreciprocal hopping. The solutions $u=-v$ and $u = -\frac{\gamma^2}{v}$ determine the critical lines in the phase diagrams that separate the regimes with bulk states localized at opposite ends; see Fig.~\ref{fig3}(c). In Fig.~\ref{fig4}(e), we present the value of $|\beta|$ as a function of $\gamma$; the stars represent the critical points with $|\beta|=1$, crossing which the bulk states will be shifted to the opposite direction. The expressions of $|\beta|$ for the dimer lattices with odd $\kappa$ are identical, so the phase diagrams are the same. In Fig.~\ref{fig4}(f), we present the spectrum of the case with $u=-0.5$, $v=1$, $\kappa=1$, and $\kappa=3$, where the critical values are also $\pm 1/\sqrt{2}$ other than $0$, same as the case with $\kappa=1$. However, for the dimer lattices with even $\kappa$, we have $|\beta|=\sqrt{|v-\gamma|/|v+\gamma|}$, so the direction of NHSE is only determined by the sign of $\gamma$. 

For the nonreciprocal dimer lattices with odd $\kappa$, the expression of $|\beta|$ is the same as the system with $\kappa=1$ [see Eq.~(\ref{betaD})], so the phase diagrams for the lattices with odd $\kappa$ are the same. On the other hand, if $\kappa$ is even, we will always have $|\beta|=\sqrt{|(v-\gamma)/(v+\gamma)|}$, so the NHSE will not depend on the value of $u$. For the dimer lattices with positive $v$, the bulk states will be localized at the left end when $\gamma>0$ and at the right end when $\gamma<0$. Thus we have fully explained the phase transition of NHSE in the dimer lattices.

\begin{figure}[t]
	\includegraphics[width=3.4in]{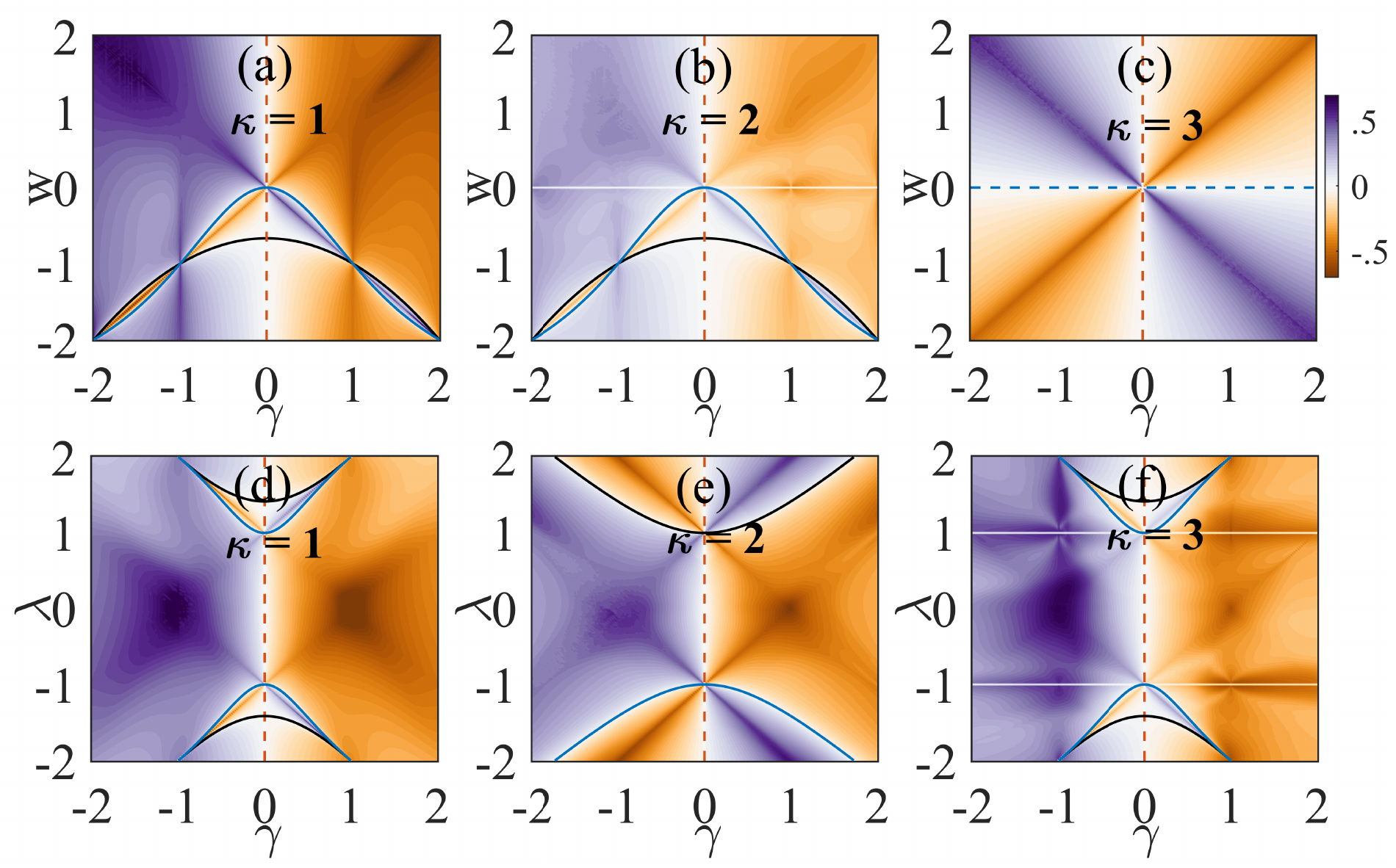}
	\caption{(Color online) (a)-(c) Phase diagrams for the mosaic nonreciprocal trimer lattices as a function of $w$ and $\gamma$. The intracell hoppings are $u=1$, $v=2$. (d)-(f) Phase diagrams for the off-diagonal nonreciprocal AAH model with $\alpha=1/4$ and $t=1$ as a function of $\lambda$ and $\gamma$. The color bar indicates the dMIPR value of the eigenstates.}
	\label{fig5}
\end{figure}

\section{Generalization to nonreciprocal trimer lattices and off-diagonal AAH models}\label{sec5}
The above discussions can be further generalized to other 1D lattices. For instance, we can study the NHSE in the mosaic nonreciprocal trimer lattices. The unit cell of the Hermitian trimer lattice contains three sites with intracell hopping being $u$ and $v$, and with the intercell hopping being $w$~\cite{Alvarez2019PRA}. By adding nonreciprocity at equally spaced sites, we can obtain the Hamiltonian of the mosaic nonreciprocal trimer lattice as
\begin{align}\label{Ht}
	H_{trimer} =& \sum_{j=1}^{N_c} \left[ u c_{j,A}^\dagger c_{j,B} + v c_{j,B}^\dagger c_{j,C} + w c_{j,C}^\dagger c_{j+1,A} + \text{H.c.}  \right] \\ \notag
	 &+ \sum_{s} \left[ \gamma c_{s\kappa}^\dagger c_{s\kappa+1} - \gamma c_{s\kappa+1}^\dagger c_{s\kappa} \right].
\end{align}
The phase diagrams are presented in Figs.~\ref{fig5}(a)-\ref{fig5}(c). For the trimer lattices with $mod(\kappa,3)=1$ and $mod(\kappa,3)=2$, we have
\begin{equation}
	|\beta|= \sqrt{\left| \frac{(u-\gamma)(v-\gamma)(w-\gamma)}{(u+\gamma)(v+\gamma)(w+\gamma)} \right|}.
\end{equation} 
Solving the equation $|\beta|=1$ gives us the following three solutions:
\begin{equation}
	\gamma = 0; \qquad w = -\frac{uv+\gamma^2}{u+v}; \qquad w = -\frac{(u+v)\gamma^2}{uv+\gamma^2}.
\end{equation}
For the trimer lattices with $\kappa$ being multiples of $3$, i.e., $mod(\kappa,3)=0$, we have
\begin{equation}
	|\beta|=\sqrt{\left| \frac{w-\gamma}{w+\gamma} \right|},
\end{equation}
which gives us two solutions: $\gamma=0$ and $w=0$. Then whether the bulk states are localized at the right or left end of the lattice is determined by the sign of $w$ and $\gamma$. For the trimer lattice with $\kappa=3$, the phase diagram is the same as the normal HN model. The phase diagram is determined by whether $\kappa$ is a multiple of $3$ or not.

We also study the commensurate off-diagonal Aubry-Andr\'e-Harper (AAH) model where the hopping amplitudes are modulated by the function $t_j=t+\lambda \cos (2\pi \alpha j)$ with $\alpha=p/q$ and $p$, $q$ being coprime integers~\cite{Ganeshan2013PRL}. The model Hamiltonian of the mosaic nonreciprocal AAH model is 
\begin{align}\label{HA}
	H_{AAH} = \sum_{j=1}^{L-1} \left\{ \left[ t+\lambda \cos (2\pi \alpha j) \right] c_j^\dagger c_{j+1} + \text{H.c.} \right\} \\ \notag
	+ \sum_{s} \left[ \gamma c_{s\kappa}^\dagger c_{s\kappa+1} - \gamma c_{s\kappa+1}^\dagger c_{s\kappa} \right].
\end{align}
By introducing nonreciprocity, we will get NHSE. We take $\alpha=1/4$ as an example here. When $mod(\kappa,4)=1$ and $mod(\kappa,4)=3$, we have
\begin{equation}
	|\beta|=\sqrt{\left|  \frac{(t-\gamma)(t-\lambda-\gamma)(t-\gamma)(t+\lambda-\gamma)}{(t+\gamma)(t-\lambda+\gamma)(t+\gamma)(t+\lambda+\gamma)} \right|}.
\end{equation}
Solving the equation $|\beta|=1$ gives us the following solutions:
\begin{equation}
	\begin{split}
		&\gamma=0; \qquad \lambda = \pm \sqrt{2(t^2+\gamma^2)}; \\
		&\qquad \lambda=\pm \sqrt{\frac{(t+\gamma)^4+(t-\gamma)^4}{2(t^2+\gamma^2)}},
	\end{split}
\end{equation}
which are the critical lines separating the regimes with NHSE at opposite direction. For the cases with $mod(\kappa,4)=2$, we have
\begin{equation}
	|\beta|=\sqrt{\left| \frac{(t-\lambda-\gamma)(t+\lambda-\gamma)}{(t-\lambda+\gamma)(t+\lambda+\gamma)} \right|}.
\end{equation}
Solving the equation $|\beta|=1$ gives us 
\begin{equation}
	\gamma=0; \qquad \lambda = \pm \sqrt{t^2 + \gamma^2}.
\end{equation}
If $\kappa$ is a multiple of $4$, then we have
\begin{equation}
	|\beta|=\sqrt{\left| \frac{t+\lambda-\gamma}{t+\lambda+\gamma} \right|},
\end{equation}
and solving the equation $|\beta|=1$ leads to
\begin{equation}
	\gamma=0; \qquad \lambda = -t.
\end{equation} 
However, if $\lambda=-t$, the lattice will be disconnected at certain sites, so we will preclude this solution. So the phase transition will only happen at zero nonreciprocity when $\kappa$ is a multiple of $4$. The critical lines and phase diagrams for the mosaic nonreciprocal off-diagonal AAH models with $\alpha=1/4$ are presented in Figs.~\ref{fig5}(d)-\ref{fig5}(f) for $\kappa=1$, $2$, and $3$, respectively.

In conclusion, when the period for the mosaic nonreciprocity (i.e. $\kappa$) is not a multiple of that for the underlying reciprocal hopping, the phase transition in the NHSE will happen at zero as well as nonzero nonreciprocity. However, if $\kappa$ is a multiple of the period for the reciprocal hopping, then the phase transition will only happen at zero nonreciprocity.

\section{Summary}\label{sec6}
In summary, we prove the existence of entirely real or imaginary spectra in the nearest-neighboring lattices with real nonreciprocal hopping amplitudes. By introducing the directional inverse participation ratio (dIPR), we have studied the phase transitions of non-Hermitian skin effect in the 1D mosaic nonreciprocal lattices and determined the critical boundaries and phase diagrams by using the non-Bloch band theory. As to the experimental realization, our model is realizable in electrical circuits~\cite{Schindler2011PRA,Luo2018arxiv,Lee2018ComPhy,Helbig2020NatPhys,Hofmann2020PRR,Zeng2020PRB} or cold atom platform~\cite{Li2020PRL,Liang2022arxiv}, where the findings reported in this work can be verified. Our work further reveals the interesting spectral properties and the phase transition of non-Hermitian skin effect in 1D nonreciprocal lattices.

\section*{Acknowledgments}
This work is supported by the Open Research Fund Program of the State Key Laboratory of Low-Dimensional Quantum Physics under Grant No. KF202109. Q. B. Z. is
supported by R\&D Program of Beijing Municipal Education Commission under Grant No. KM202210028017. R. L. is supported by NSFC under Grant No. 11874234 and the National Key Research and Development Program of China (Grant No. 2018YFA0306504).

%\section*{Appendix}
%\setcounter{equation}{0}
%\renewcommand{\theequation}{{A}\arabic{equation}}

\end{document}